\def\year{2021}\relax
\documentclass[letterpaper]{article} % DO NOT CHANGE THIS
\usepackage{aaai21}  % DO NOT CHANGE THIS
\usepackage{times}  % DO NOT CHANGE THIS
\usepackage{helvet} % DO NOT CHANGE THIS
\usepackage{courier}  % DO NOT CHANGE THIS
\usepackage[hyphens]{url}  % DO NOT CHANGE THIS
\usepackage{graphicx} % DO NOT CHANGE THIS
\usepackage{ulem}
\usepackage{natbib}  % DO NOT CHANGE THIS AND DO NOT ADD ANY OPTIONS TO IT
\usepackage{caption} % DO NOT CHANGE THIS AND DO NOT ADD ANY OPTIONS TO IT
\usepackage[utf8]{inputenc} % allow utf-8 input
\usepackage[T1]{fontenc}    % use 8-bit T1 fonts
\usepackage{hyperref}       % hyperlinks
\usepackage{booktabs}       % professional-quality tables
\usepackage{amsfonts}       % blackboard math symbols
\usepackage{amsthm}
\usepackage{amsmath}
\usepackage{nicefrac}       % compact symbols for 1/2, etc.
\usepackage{microtype}      % microtypography
\usepackage{adjustbox}
\usepackage{bbm}            % indicator function
\usepackage{mathtools}
\usepackage{multirow,tabularx}
\usepackage{enumitem}
\usepackage[utf8]{inputenc}
\usepackage[english]{babel}
\usepackage{comment}
\usepackage[htt]{hyphenat}

\renewcommand{\paragraph}[1]{\vspace{0.3\baselineskip}\noindent\textbf{#1}}

\newcommand{\ignore}[1]{}
\newcommand{\nop}[1]{}
\newcommand{\eg}{{\textit{e.g.}}}

\newcommand{\ie}{{\textit{i.e.}}}

\usepackage{pgfplots}
\usetikzlibrary{pgfplots.groupplots}
\usetikzlibrary{shapes.geometric}
\usetikzlibrary{positioning,fit,shapes.geometric,backgrounds, calc}
\tikzset{textnode/.style={inner sep=0pt,outer sep=0,execute at begin node={\strut}}}
\tikzstyle{state} = [textnode,circle, draw, inner sep=0pt, outer sep=0]

\usepackage{lipsum}

\usetikzlibrary{calc}
\usetikzlibrary{arrows,decorations.markings}
\usetikzlibrary{shapes.arrows}
\usetikzlibrary{patterns}
\selectcolormodel{cmyk}
\pgfplotsset{compat=1.14}

\usetikzlibrary{shapes,arrows, arrows.meta}
\tikzstyle{box} = [rectangle, thick, rounded corners=2pt, outer sep=3pt, inner sep=0pt, minimum height=12pt, minimum width=48pt, text opacity=1]
\tikzstyle{bluebox} = [draw=blue, fill=blue, fill opacity=.05, box]
\tikzstyle{bluebox2} = [draw=blue, fill=blue, fill opacity=.2, box]
\tikzstyle{orangebox} = [draw=orange, fill=orange!95!black, fill opacity=.1, box]
\tikzstyle{brownbox} = [draw=brown, fill=brown!80!black, fill opacity=.1, box]
\tikzstyle{redbox} = [draw=red, fill=red!10, box]
\tikzstyle{dottedbox} = [box, dotted, ultra thick, draw=red, line width=2pt, minimum width=47pt, minimum height=30pt, text opacity=1]

\tikzstyle{edge} = [shorten >=1pt, auto, line width=0.2pt]
\tikzstyle{bluearrow} = [-Triangle, line width=1pt, draw=blue!80!black, fill=blue!70]
\tikzstyle{orangearrow} = [-Triangle, line width=1pt, draw=orange, fill=orange!80]
\tikzstyle{redarrow} = [-Triangle, line width=1pt, draw=red, fill=red!80]
\tikzstyle{greenarrow} = [-Triangle, line width=1pt, draw=green!80!black, fill=green!80!black]
\tikzstyle{blackarrow} = [-Triangle, line width=1pt, draw=black, fill=black, dashed]

\title{HetSeq: Distributed GPU Training on Heterogeneous Infrastructure}
\author {
    Yifan Ding,
    Nicholas Botzer, 
    Tim Weninger \\
}

\affiliations{
Department of Computer Science \& Engineering \\ 
University of Notre Dame \\ 
Notre Dame, IN, USA \\
    \texttt{\{yding4,nbotzer,tweninge\}@nd.edu} 
}
\begin{document}

\maketitle
\begin{abstract}
Modern deep learning systems like PyTorch and Tensorflow are able to train enormous models with billions (or trillions) of parameters on a distributed infrastructure. These systems require that the internal nodes have the same memory capacity and compute performance. Unfortunately, most organizations, especially universities, have a piecemeal approach to purchasing computer systems resulting in a heterogeneous infrastructure, which cannot be used to compute large models. The present work describes HetSeq, a software package adapted from the popular PyTorch package that provides the capability to train large neural network models on heterogeneous infrastructure. Experiments with language translation, text and image classification shows that HetSeq scales over heterogeneous systems. Additional information, support documents, source code are publicly available at \href{https://github.com/yifding/hetseq}{https://github.com/yifding/hetseq}.
\end{abstract}

\renewcommand{\sout}{\vphantom}

\section{Introduction}
\label{sec:introduction}

The AI community has witnessed rapid growth in the number of neural network models. Many of these models have matched or even exceeded human performance in a wide range of areas including image classification~\cite{simonyan2014very, he2016deep},
natural language processing~\cite{vaswani2017attention, devlin2018bert, peters2018deep, radford2019language, yang2019xlnet}, 
%cite: tranformer, bert, emo, gpt2, xnet
and multiplayer online battle arena (MOBA) games like DOTA2~\cite{berner2019dota}, and StarCraft II~\cite{vinyals2019alphastar}.

%cite: open5 and deepstar/openstar

These increasingly powerful neural network models operate over extremely large training data and require millions (or even billions) of model parameters~\cite{pudipeddi2020training}. This requires enormous computational resources that are only available to a handful of large organizations. Alongside the investment in hardware including CPU, GPU and high performance file systems, economies of scale present in only the largest organizations allow for the management and rapid development of new technologies like cloud computing (\eg, AWS, Azure), tensor processing units (TPUs), half-precision computation (\eg, Volta or Turing architectures of NVIDIA GPUs), and network architectures (\eg, infiniband) for high performance communication between thousands of these systems. 

%Furthermore, popular deep learning software platforms like Tensorflow and PyTorch, developed at large technology companies, assume that models will be trained on one or more \textit{identical} nodes.

Smaller organizations like startups and universities, on the other hand, have relatively limited resources and typically purchase computing systems in an ad hoc manner -- whenever funds allow. University computing resources are therefore much more heterogeneous in their composition compared to large technology companies and government labs. Furthermore, mismatches in memory capacity, network interface, and GPU/CPU capacity limit the development of large models. 
Given the system-homogeneity assumptions of deep learning software platforms, even the process of training an existing model on new data is difficult or impossible because the resource limits of a single node in a heterogeneous system limits the entire pipeline. With these limitations, training even moderately sized models could take weeks or even months to complete. 

\vspace{.2cm}
\noindent If not corrected, universities and other small organizations risk losing relevance in the race to develop newer and better machine learning models.%~\cite{strubell2019energy}.
\vspace{.2cm}

%TODO:
%{\color{Red} Rewrite the motivation, 0. mention current limit.
%1. mention challenges: a.dataset, loader. b. ckpt.  c. hardware limitation. (memory storage, memory bandwidth, shared hardware, )  2. mention our methods 3. talk about the significance of our implementation: experiments. use on the fly, speed up, scalability, preserve performance  4. other observation: smaller model, memory bandwidth,  }

In the present work, we endeavour to level the playing field by adapting existing software platforms to train large neural network models on heterogeneous systems. We release our software package called HetSeq at \href{https://github.com/yifding/hetseq}{https://github.com/yifding/hetseq}, which is built on PyTorch and includes the common GPU environment and NVIDIA Collective Communications Library (NCCL) without any extra libraries or packages. In contrast, most existing distributed GPU training platforms~\cite{sergeev2018horovod,paszke2019pytorch} require extra packages like Docker and Open MPI, which may not be deployable over shared file system without administrative privileges. 
%HetSeq can be run directly from the command line or it can be integrated with popular task queue systems. HetSeq also manages initialization, randomization, and GPU assignment, which can be arduous on heterogeneous multi-node and multi-GPU computing clusters. 
%As a result, HetSeq can be directly used to train most large supervised learning models on nearly any collection of networked computing hardware. %Our BERT model also provides an example of just-in-time (JIT) data loading to avoid memory exceeding in large datasets . 

We evaluate the computational performance of HetSeq using NVIDIA GPU nodes with various number of cores, memory capacity, and CPU architecture, while running in competition with the myriad of other machine learning projects at the University of Notre Dame. We perform  experiments on typical translation, language modelling, and image classification tasks. We show vast improvements in training scalability using HetSeq without sacrificing model performance.
Finally, the released project code covers detailed steps to install and execute. Examples of language translation, language modeling, and image classification tasks are included. Furthermore, developers and researchers can easily extend HetSeq to many other models with little effort.

% so that the code can be easily adapted to other models.  punchline...

%Under this limit, we want to train large models by fully utilizing the resources such that we can reduce the gap to push research forward. Here we describe a general data parallel method to train large models faster on heterogeneous systems (i.e. GPU work stations). Basically, we train a model by splitting training data and distribute different data segments to different GPUs. Each GPU has its individual model and optimizer. Different GPUs communicate about model parameters in feedforward stage and losses just before backpropagation. GPU reduces the training time by performing backpropagation in parallel. Our methods achieve up to $7$ times faster (in 8 GPU work stations with 32 GPUs) comparing to single GPU work station (with 4 GPUs). We observe that relatively smaller model (still as large as millions of parameters) achieves better performance especially scalability. We evaluate our method on language model BERT and translation model Transformer. Our method is highly flexible and can be easily applied to many other large models. We will release our code in the future. 

\section{Preliminaries}
\label{sec:preliminaries}
% Standard training archs look like this. describe tensorflow, pytorch Keras
% They have these pieces
%optimizer, model, how they all fit together
Many deep neural network (DNN) models are built upon popular platforms like TensorFlow~\cite{abadi2016tensorflow}, PyTorch~\cite{paszke2019pytorch} and Apache TVM~\cite{chen2018tvm}. A standard DNN model with backpropagation includes: 
\vspace{-.1cm}
\begin{enumerate}
    \item Model
    \item Dataset and Dataloader
    \item Optimizer and Learning Rate Scheduler
    \item Checkpointing
\end{enumerate}
\vspace{-.1cm}
A model can be described as a directed graph, where nodes represent parameters and edges represent the dependencies of the pipeline. A dataset is defined as the actual input of the model and dataloader executes the data loading process from disk into memory. The optimizer plays as a key role by updating the model parameters according to calculated gradients and learning rate generated by scheduler. Finally, checkpointing is often used to store and load training snapshots.

In the typical case, after the model is defined and its parameters are initialized or loaded from a previous checkpoint, the indices of the dataset are loaded into memory. Because modern datasets are usually too large to fit into GPU memory all at once, the dataloader constructs smaller subsets of the dataset called batches to pass to the model one by one. Upon receiving each batch of data, the model performs a forward pass of the data over the model parameters and computes a loss function. Based on the results of the loss function, learning gradients are obtained by performing backpropagation. For each parameter, a pre-defined optimizer 
%computes a partial derivative with respect to object function,
takes its gradient, 
the corresponding learning rate from learning rate scheduler, and other required factors to update the parameter. For a single batch of training data, this whole process including the forward pass, backpropagation, and the parameter update is called one step. One epoch is complete when all the batches have been processed over the entire training data. 

%Finally, once the training steps/epochs reach a certain threshold or another defined training criteria (\eg, loss) is reached, we can store a checkpoint. The checkpoint stores the model parameters as well as necessary training status like optimizer states and learning rate. 

% do you think we should include more details for the checkpoints or stopping criteria?

% i'm not sure. I just thought the way that sentence was worded was to long and had to many things in parentheses going on. It was a lot to understand.

Finally, once the training steps/epochs reach a certain threshold or another defined training criteria is reached (based on the objective function, learning rate, etc), we store a checkpoint. The checkpoint stores the model parameters as well as other necessary training status (like optimizer status, learning rate status etc.)

%nice... I like this change ^^

%3 parallelism: model, pipeline, data
\paragraph{Parallel Processing.} The model training process is costly, especially using traditional CPU processing. Fortunately, most deep learning platforms support highly parallel GPU processing as well.
There are three popular GPU parallelism: model parallelism~\cite{shoeybi2019megatron, shazeer2018mesh}, pipeline parallelism~\cite{harlap2018pipedream, huang2019gpipe} and data parallelism~\cite{tarditi2006accelerator}. 
Model parallelism refers to splitting models into several parts where different parts are distributed to different devices (GPUs). In a backpropagation schema, intermediate output from the previous device is transferred to next device in the forward step while the gradients of next device are transferred to previous device during backpropagation. Model parallelism is essential especially for very large models which cannot fit the limited memory of a single GPU. However, heavy intermediate output and gradient communication may cause high latency. 
Pipeline parallelism is similar to model parallelism, instead of splitting the model into multiple steps, pipeline parallelism splits a single step into multiple parts. 
Data parallelism approaches split the training data into different parts to be distributed into different devices. Each device has its own model, data batch and optimizer thus performing forward, backward, parameter update individually. Recently, researchers from Microsoft have developed a combined platform using aspects from data parallelism, model parallelism, and pipeline parallelism to successfully train a trillion-parameter language model~\cite{rajbhandari2019zero, pudipeddi2020training}.

\paragraph{Heterogeneous Infrastructure.}
These state-of-the-art distributed parallel frameworks work well when the infrastructures are homogeneous, having the same memory restrictions, GPU capacity and CPU throughput with high inter-node communication speed. Unfortunately, the business models of many smaller organizations necessitate the need for a more piecemeal approach to their system purchases. As a result, their computing infrastructure is heterogeneous, with many different types of systems purchased individually and without coordination. In this heterogeneous setting, the deep learning system cannot assume that each individual system, CPU, and GPU will have identical memory and throughput. Instead, communication across nodes with different networking infrastructure can be costly. Furthermore, the principles of model parallelism and pipeline parallelism cannot be easily applied because differences in memory capacity and throughput performance cause severe conflicts when reconciling the distributed computation. As a result not all models and training settings are compatible with certain systems and perform poorly on heterogeneous infrastructure.  

%{\color{red}We can extend Heterogeneous here}
%what do you want to say?
% i am thinking our title includes Heterogeneous, we need to emphasize the properties. Like communication cross nodes can be costly. Model parallelism and pipeline parallelism can not be well applied. 
% about to finish first draft of the figure 1 part 1.  <ok

%% YIFAN ^^ check out the heterogeneous infrastructure discussion above

The HetSeq package, described in the present work, is a distributed deep learning package that provides the capability to train large models on heterogeneous distributed infrastructure. HetSeq is adapted from the fairseq subpackage within PyTorch and uses principles of data parallelism to avoid heavy data communication so that each GPU process can execute expensive forward and backward passes and parameter updates locally. Other inter-node communication of the training loss, gradients, and parameters is managed carefully.

%We select data parallelism as our method to avoid relatively heavy data communication. So each process (GPU) execute expensive steps including forward and backward propagation, and parameter updates locally.

%At each step, a mini-batch is divided evenly across all the data parallel processes, such that each process executes the forward and backward propagation on a different subset of data samples, and uses averaged gradients across processes to update the model locally

\section{HetSeq: Distributed GPU training on Heterogeneous Systems}
\label{sec:distributed}
% Describe each of our pieces - probably keeping same outline as last section

\begin{figure*}[ht]
    \centering
    \includegraphics[width=\linewidth]{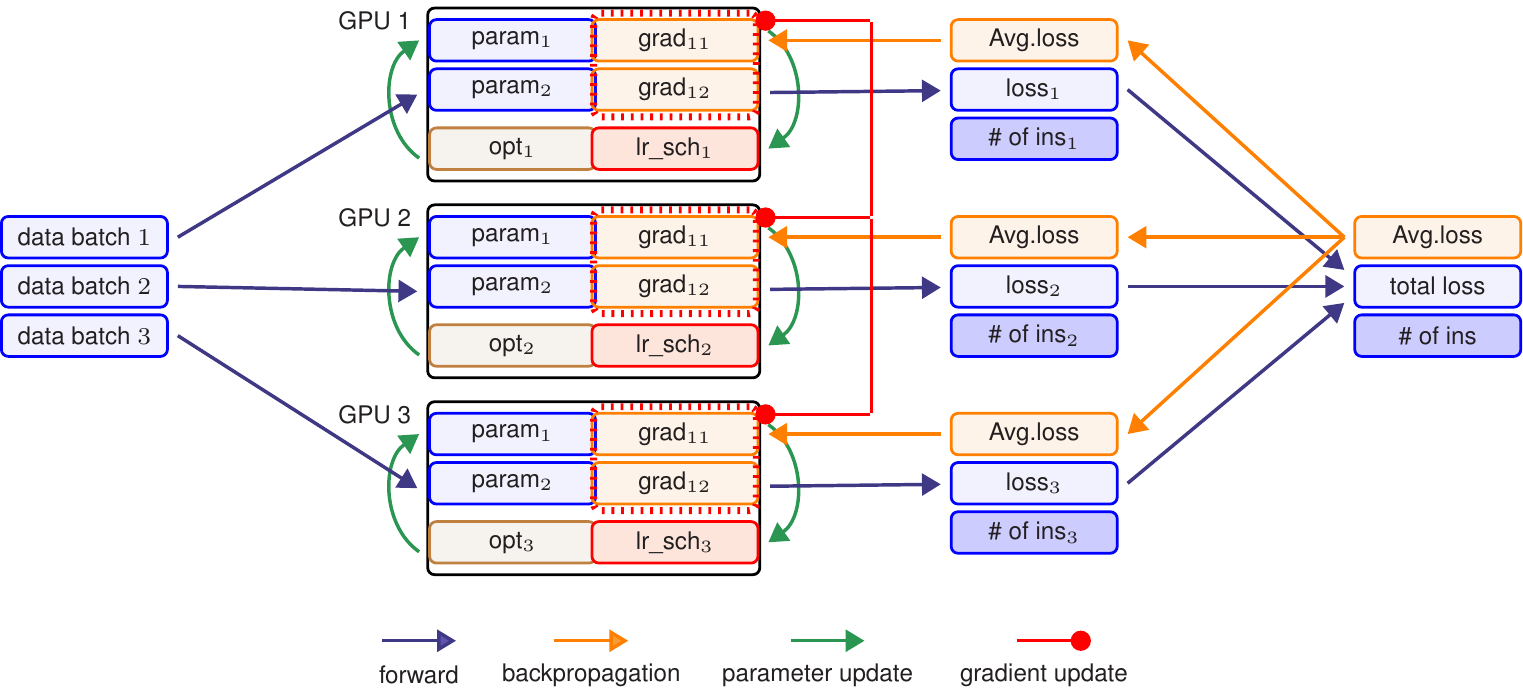}
    \caption{Distributed Data Parallel pipeline employed in HetSeq. The training can be divided into four major steps: the forward step, backpropagation, gradient update, and parameter update. Each GPU has individual models, optimizer (opt) and learning rate scheduler (lr\_sch). Model, optimizer, and learning rate scheduler are all initialized with the same states while different GPUs process different segments of training data.
    %In the forward pass, data batches that maximize the available memory on each constituent GPU  are loaded to different GPUs 
    In the forward pass, data batches assigned to different GPUs are loaded, the forward pass is executed and individual loss functions are computed. Different loss functions from different GPUs are aggregated to compute the average loss. In the backpropagation step, 
    the average loss is distributed back to each GPU to calculate individual gradients. Gradient updates are then applied to communicate the gradients back to each GPU. Finally, the optimizer and learning rate scheduler of each GPU performs parameter updates individually. 
    }
    \label{fig:pipeline}
    \vspace{.5cm}
\end{figure*}

%Compared to CPUs, deep neural network training can be orders of magnitude faster on a GPU. This is because the GPU have specialized logic for the vector and matrix calculations commonly used in neural networks, but also because they are able to process more input training instances (\ie, handle larger a batch size) in parallel. 

{\color{red} \sout{Compared to CPUs, GPU nodes are expensive and have restrictive memory capacity. Because of the memory bottleneck especially, a lot of effort has been made to make the best use of the GPU memory by optimizing the learning process~\cite{rhu2016vdnn, chen2018modnn}. However, with very large models, it is common to find that a single GPU might only fit a handful of training instances (sometimes fewer than ten!). So, as described in the preliminary section above, it is common to integrate multiple GPUs across multiple nodes to increase the batch size. For example, if we use four GPUs on a single node and each GPU can run a batch with 8 instances, then a total batch size of 32 can be achieved. If we scale to 32 GPUs on 4 nodes, the total batch size is therefore scaled to be 32$\times$8=256. If we assume no communication overhead and 100\% parallelism, this integration will increase training speed for a factor of $8$! Unfortunately, we cannot assume 100\% parallelism; therefore communication latency, especially heterogeneous systems, quickly degrade performance.}}

%However, we have {\color{red} the limitation is the GPU memory capacity. With large model, big data, we can integrate multiple GPUs to meet the requirement of the memory: For large model, we need to increase the batch size to speed up the training. However, larger batch size requires larger memory, which can be achieved by multiple GPUs. Besides, we have communication problem. The core idea of the parallelism is to reduce communication and latency. Different GPUs need to communicate loss. Ideally, different GPUs should get the same gradient, and after get gradient, individual GPU can update the parameters. Once the parameters are updates, different GPUs can communicate with each other for synchronization.
%Different GPUs communicate about model parameters in feedforward stage and losses just before backpropagation. GPU reduces the training time by performing backpropagation in parallel.}

%{\color{red}We can extend Heterogeneous here}

HetSeq is designed to perform distributed {data parallel} GPU training across heterogeneous systems with large models and training datasets. {\color{red} \sout{To do so HetSeq revised several parts of the typical training process to be better compatible to a wide range of large model and datasets.}} Specifically, we modified modules to enable fast communication and avoid structural hazards like {\color{red} \sout{CPU and}} exceeding memory capacity. In the HetSeq system implementation, each GPU contains an individual process and inter-process communication (IPC) is utilized.  HetSeq is adapted from PyTorch
\footnote{\hyperlink{https://github.com/pytorch/pytorch}{https://github.com/pytorch/pytorch}}, especially the DistributedDataParallel (DDP) mechanism\footnote{\hyperlink{https://pytorch.org/docs/stable/notes/ddp.html}{https://pytorch.org/docs/stable/notes/ddp.html}},
and fairseq 
\footnote{\hyperlink{https://github.com/pytorch/fairseq}{https://github.com/pytorch/fairseq}}. 

As shown in Figure~\ref{fig:pipeline}, different GPUs perform forward pass, backpropagation, and updating parameters in parallel. Individual GPU processes also communicate parameters, gradients, and loss functions. In the remainder of this section we introduce main aspects of HetSeq one by one and describe how they handle the complications that arise in the heterogeneous setting.

\subsection{Model Initialization}

%%YIFAN - what is wrong with the current approach? <<<<

%% the initial idea I develop this is that I ask many people in our department and no one has used distributed training. Then I search for some online resources, they are either not releasing the training code or have strict form to follow which makes it hard to apply to one's research code. 
%% the ddp itself is available on pytorch and very few people except big company is using it. What we want to achieve here is an open package one can easily use for his own application.

%% Got it! i understand

% Model initialization
After the \texttt{ProcessGroup} is initialized, HetSeq initializes the model. A model is defined as a child class of \texttt{torch.nn.module}, which takes an input tensor (\ie, images or sentences), and outputs a real number from the loss function. This model initialization is performed once on the master node and broadcast to all other GPUs so that they share the same initial state.

\subsection{Dataset and Dataloader}
% Core: very large inputs, multiple input tensors with different shapes and lengths, enough shuffle and results can be repeated. Allow multithreading and multiprocessing. 
The format of the input data varies widely depending on the application. Data access should support multiprocessing and multithreading, and the data access medium should support shuffling of the training instances in a way that can be easily stored and reproducible. 

Unlike the typical training process on a single GPU, distributed training on multiple nodes with multiple GPUs has many challenges. When the dataset is small, then the system should just load it from disk into memory, and in each training step a chunk of the dataset passed to the GPU. However, this is not feasible for even medium-sized datasets. In these typical cases, the dataloader is a bottleneck in training large models. Our solution is to separate the dataset into shards so that the dataset can be loaded in parallel. In addition to the dataset size and dataloader speed, the index sampler must be able to accommodate multiple dependent tensors with different data types. Simply put, our goal is to find a universal input and output mechanism that can quickly handle arbitrary, dependent tensors stored over multiple shards.

For this HetSeq uses HDF5 wrapped by h5py\footnote{\hyperlink{https://github.com/h5py/h5py}{https://github.com/h5py/h5py}} packages as our main strategy to deal with the dataloading challenges. HDF5 supports self-describing and heterogeneous data at scale. Another benefit is that it can group multiple relative tensors together in a hierarchical manner so that it can be loaded faster with multiple processes and threads. We define our dataset class as a child module of $\texttt{torch.utils.data.Dataset}$. $\texttt{\_\_len\_\_}$ and $\texttt{\_\_getitem\_\_}$ functions must be implemented in the model definition, and the file $\texttt{open}$ function must be defined inside the $\texttt{\_\_len\_\_}$ and $\texttt{\_\_getitem\_\_}$ functions instead of the $\texttt{\_\_init\_\_}$ function to support multithreading loading in the PyTorch dataloader. In order to handle data shards, we add another class to accumulate the lengths of each file. The index of each training instance is mapped to an offset location at a corresponding shard.

\subsection{Forward Pass}
%trick of communicate object function.
Having obtained the lengths and indexes of all shards, we generate sampler indices by forming batches that satisfy some criteria like maximizing the number of instances in a batch (\ie, batch size) or maximum number the of tokens in a batch. In a distributed training forward pass, each GPU has separate sampling indices according to its GPU index. Each GPU has an individual dataloader to load corresponding data by looking up sampling indices to form a batch. Once a data batch is ready, each GPU can immediately complete the forward pass and compute its individual loss function.  

\subsection{Optimizer and Learning Rate Scheduler}
Each GPU has its own optimizer as well as a learning rate scheduler. Both are initialized with the same parameters. \sout{In a typical optimizer, parameters are updated by gradient decent with a fixed learning rate.}  Different from the backpropagation process in a homogeneous system, heterogeneous GPUs, with different memory capacities, may require different batch sizes or a different number of tokens. So the loss functions of individual GPUs will likely have different weights. 

During the last step of an epoch, GPUs may contain partially-filled batches and even empty batches. For example, if there are 5 total training instances and the batch size is set to 2 globally and we want to perform distributed training on 4 GPUs named A, B, C, and D, then the corresponding batch sizes should be 2, 2, 1, and 0 respectively, where batch C is half-filled (1/2) and batch D is empty (0/2). If we take the average or sum of the loss directly, then we will not compute the average loss in a way equivalent to the non-distributed setting. Instead, we augment the output by associating it with weights like batch size or number of tokens. After all GPUs send their output, we use a weighted sum to obtain the average. This task is performed by the master process. When complete, the master broadcasts the average loss to all the other processes. 

% backward pass and optimization step
\subsection{Backpropagation}
As soon as a GPU receives the average loss from the master, it can perform backpropagation to obtain the gradient for its individual model. However, because individual models stored on each GPU consider different data (in parallel), their parameters are likely to diverge. It is important to ensure that each parameter has the same partial derivative across GPUs. The $\texttt{DistributedDataParallel}$ (DDP) class of PyTorch supports partial backpropagation and gradient synchronizations across GPUs; however, if one GPU has empty batch, we provide the GPU a dummy batch by copying its very first data batch and setting the gradient to $0$ before backpropagation. %Returning to the previous example where the size of batch$_D$ was 0, our solution in the first step can cause error because its first data batch is also empty. Theoretically, it can cause error when the number of batches is less than the number of GPUs. But in reality, it is not a problem since the number of batches is way more than thousands. 
After the backpropagation pass is complete, the calculated gradient is broadcast to each GPU. Then the GPU's optimizer retrieves the learning rate from the a learning rate scheduler and updates the parameters in the model. 

\subsection{Checkpointing}
%\sout{Storing checkpoints is a key component of the training process. Storing a model's architecture and associated parameters enables performance evaluation and fine-tuning. Checkpointing also enables large models to be trained over different phases that alter the batch size, optimizer, learning rate scheduler, etc. We can also stop and continue training at any checkpoint as well as record the training process by storing multiple checkpoints over time. But heterogeneous infrastructures provide additional challenges to the standard checkpointing procedure because the current state of the model is tightly integrated with the capacity and capability of each node.} 
In HetSeq, the master process is responsible for loading and storing checkpoints. In addition to include model parameters in checkpoints, we also need to consider: (1) the number of completed epochs, (2) the number of completed steps, (3) the optimizer status (including the learning rate scheduler status), and (4) the random number seed, as well as several other settings. %We will briefly talk about how to continue training exactly from a training checkpoint using exactly the same training settings. For other changes, one can redefine part of the settings and replace the old settings before new training. 

%{\color{red} \sout{When a checkpoint is loaded without any manual changes several steps must occur. First and foremost, we load the model parameters from the checkpoint. We also load the optimizer parameters, history, and other related information, as well as the learning rate scheduler status. Complications arise, however, when reconstructing the dataset and dataloader. Specifically, we need to know exactly how the training data was split, shuffled, and assigned to each GPU at the time of the checkpoint. To resolve these complications, HetSeq keeps a random number controller. This random number controller carries a single seed, which is stored with the checkpoint, as well offsets for each random process. Specifically, given a random number seed $S$, we initialize all the random seeds of all package dependencies (\eg, pytorch, numpy) to $S$. Whenever a random process is needed, we set the seed to be $S + offset$ and revert the seed back to $S$ afterwards. For example, when performing the shuffle and batch building process in epoch number $N$ we use the seed $S + N$. For other random processes in some step $P$, use use the seed $S + N + P$. }
%\sout{In this manner, we can load the training process to the exact same state as when the checkpoint was taken.
%}
%}

%\iffalse
\subsection{Additional Considerations}
\subsubsection{Delayed Update (Gradient Accumulation)}
Compared to the forward pass, backpropagation is a relatively expensive process. It needs to compute the gradient for the whole model and update each parameter. Because of the difference in complexity, PyTorch recently implemented delayed updating, which aggregates the loss function computed from multiple forward passes before performing the backpropagation pass~\cite{ott2018scaling,youkawa2018delayed}. However, when using delayed update the batch size is essentially scaled by the number of forward passes. Changing the batch size and data splitting is further complicated in the heterogeneous infrastructure because batch sizes may be different according to each node's capacity. %For example, if we implement a batch size of 8 on 8 GPUs, then the data splitting is equivalent to performing a delayed update of 8 forward passes.
Because batch sizes dramatically influence training performance, we need to consider the scaling effect of delayed update when computing the average loss function. In addition, carefully managing delayed update settings is important for reproducing model results. 

%Like how batch sizes are computing to maximize each node's capacity, we also need to carefully %But Similar to scale batch size by distribute data across GPUs, we can also scale batch size by aggregating multiple forward passes. The object function and associated counters are accumulated within multiple forward pass. After the number of forward pass exceeds certain threshold, forward pass is stopped and backward pass starts to execute. 
%Delay update reduces the latency in the backward pass and also help us test our distributed settings. 

%For example, if we implement batch size = $8$ on each GPU and utilize $8$ GPUs in total. The data split setting is equivalent to perform delay updates with update frequency = $8$ on a single GPU with $8$ batch. We can observe initial loss function curve to adjust our optimizer and learning rate scheduler to obtain better performances.

%so the point for us is that delayed updating is great, it improves the performance, but it messes with how we calculate the batch size, which is important in the heterogeneous setting because it makes the data splitting complicated. << right?
% yup, and if some one wants to reproduce some experiments, have the correct batch size is very important. In our experiment, we also show that keep the same setting while only changing the batch size to a large number, the performance can be much worse.  PERFECT

%2. test training setting on distributed can be difficult, can be simply simulated by employing delay update on a single GPU. 

\subsubsection{Cython}
Compared to C/C++, Python is much slower. Performance differences are exacerbated when looping over large datasets and shuffling billions of training instances. HetSeq uses Cython and C++ bindings whenever possible for quicker runtime. %one can replace a function in python with much faster running speed in C/C++. It is widely used to evaluate translation with GLUE score and other loop based expensive computation.
 
\subsubsection{Prefetch and Cache}
Even though HetSeq achieves multi-processing and multi-threading on heterogeneous infrastructure, data loading is still a bottleneck before each forward pass, especially on larger batch sizes. We use prefetching and caching to reduce data loading latency. With prefetch, instead of loading every batch just before the forward pass, we fetch the next batch while training on the current batch. When memory capacity allows we can prefetch multiple batches. For caching, we utilize the least recently used (LRU) policy to store data in the memory. Although the LRU cache saves some disk access, the use of prefetch results in considerable performance gains because of the size of most datasets.
%\fi

\section{Extending the HetSeq Package}

\begin{figure}
    \centering
    \includegraphics[width=\linewidth]{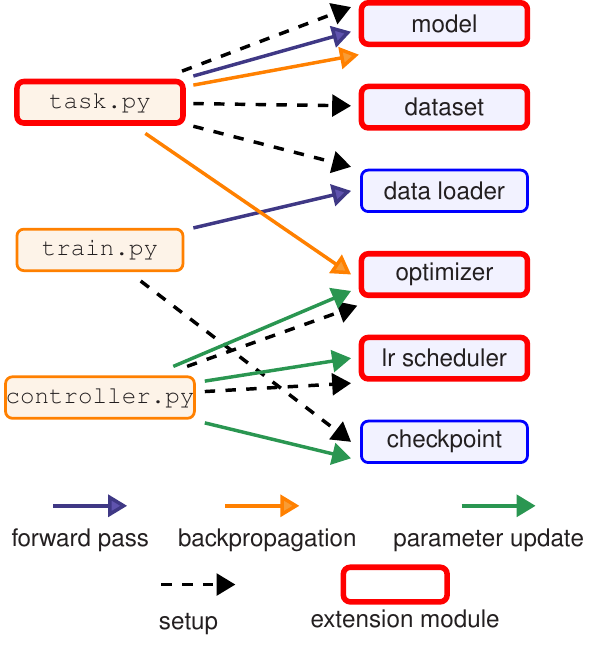}
    \caption{Module and component architecture of the HetSeq package. Python modules on the left are responsible for various components listed on the right. Modules and components highlighted in red can be extended to perform other machine learning tasks over heterogeneous infrastructure.}
    \label{fig:hetseq}
    % \vspace{-.5cm}
\end{figure}

The HetSeq package contains three major modules illustrated on the left in Figure~\ref{fig:hetseq}: train.py, task.py, and controller.py to coordinate the main components illustrated on the right.
The train.py module initializes the distributed system and its various components.
The task.py module defines the model, dataset, data loader, and optimizer functions; it also executes the forward pass and backpropagation functions. The controller.py module acts as the main training controller. It executes the actual model, optimizer, and learning rate scheduler; loads and saves the checkpoint; communicates the loss; and updates the parameters.

%All together, the three modules define the six components during the setup, forward pass, backpropagation, and parameter update steps. The model component calculates a loss from data inputs, which is organized into batches and indexed by the dataset component. When the batch assignment is done, the data loader loads the the batches to the proper devices. The optimizer and learning rate scheduler operate in tandem to execute the distributed parameter update. Finally, the checkpoint component is used to load and/or store the training state.

%2.how to use the package to run BERT and MNIST
To extend the HetSeq package to other tasks, the five components highlighted in red in Figure~\ref{fig:hetseq} need to be defined by using existing plug-ins or by defining customized components. We also provide documentation on how to install, use, and extend HetSeq which is publicly available at \href{https://hetseq.readthedocs.io}{https://hetseq.readthedocs.io}.

\section{Experiments}
\label{sec:experiments}
%HetSeq is currently focuses on sequence to sequence models, but can be easily adapted to consider a wide range of additional deep neural network models. {\color{red} We have included an example of image classification model in our package.} 
Here we test the performance of HetSeq on three popular deep neural network models: (1) the transformer translation model, (2) the BERT language model, and (3) an image classification model for MNIST handwritten digit database. We used a variety of different heterogeneous distributed setups. Described in Tab.~\ref{tab:infrastructure}, each GPU node has $4$ GPUs consisting of a various number of cores and main memory. Experiments evaluate the training speed, scalability, and model performance across various configurations, specified in Tab.~\ref{tab:exp_config}. Simply put, experiments with homogeneous regime use the same GPU nodes while heterogeneous settings use different combinations of different GPU nodes. For different heterogeneous settings, we keep the number of epochs constant while changing the number of steps on each GPU. For example, a single epoch with $16$ steps on one GPU is equivalent to one step per GPU over 16 GPUs. All the other settings are set to the same for the same task.

\begin{table}[t]
    \centering
    \caption{Infrastructure used in experiments.}
    \begin{adjustbox}{max width=\columnwidth}
    \vspace{-.3cm}
    \begin{tabular}{@{}l r r r l r r r r@{}}
    \toprule
        \multicolumn{3}{c}{\textbf{CPU}} & & \multicolumn{5}{c}{\textbf{GPU}} \\
        \cmidrule{1-3} \cmidrule{5-9}
           \textbf{Type} & \textbf{Cores} & \textbf{Mem}  & & \textbf{Type} && \textbf{\#} & \textbf{Cores} & \textbf{Mem}  \\ \midrule 
           Xenon  &16/24  & 96/128GB & & Xp && 4 & 3840    & 12GB \\ 
           Xenon  &24    & 128GB & & 1080Ti && 4 & 3584    & 11GB \\ 
           Xenon  &24    & 128GB & & P100 && 4 & 3584    & 16GB \\ 
           \bottomrule
    \end{tabular}
    \end{adjustbox}
    \label{tab:infrastructure}
    %\vspace{-.3cm}
\end{table}

%\vspace{.3cm}
\begin{table}[t]
\caption{Experiment Configurations}
\vspace{-.3cm}
\begin{center}
    \begin{adjustbox}{max width=\columnwidth}
    \begin{tabular}{@{}l l l@{}}
    \toprule
         \textbf{Nodes} & \textbf{Translation Config.} & \textbf{BERT \& MNIST Config.}\\ \midrule 
         1 & 1080Ti$\times$1 & P100$\times$1\\
         2 (hom) & P100$\times$2 & Xp$\times$2\\
         2 (het) & P100$\times$1 + Xp$\times$1 & P100$\times$1 + Xp$\times$1\\
         4 (hom) & P100$\times$4 & Xp$\times$4\\
         4 (het) & P100$\times$1 + Xp$\times$1 + 1080Ti$\times$2 & P100$\times$2 + Xp$\times$2\\
         8 (het) & P100$\times$1 + Xp$\times$4 + 1080Ti$\times$3 & P100$\times$4 + Xp$\times$4\\ \bottomrule
    \end{tabular}
    \end{adjustbox}
    \label{tab:exp_config}
\end{center}
\end{table}
\vspace{.2cm}

\begin{table*}[t!]
%\caption[dd]{Transformer results on English-to-German dataset. Homogeneous (hom) and heterogeneous (het) experiments over 1-, 2-, 4-, and 8-node node configurations shows HetSeq on heterogeneous configurations scales on the transformer translation task at about the same rate as the homogeneous configurations. On 8 heterogeneous nodes HetSeq achieve almost a 5x speedup without a significant loss in model performance.}
\caption[dd]{We evaluate HetSeq on Translation, BERT, and MNIST. Homogeneous (hom) and heterogeneous (het) experiments over 1-, 2-, 4-, and 8-node node configurations. HetSeq on heterogeneous configurations scales on the translation task at about the same rates as the homogeneous configurations. On 8 heterogeneous nodes, HetSeq achieves almost a 5x speedup on the Translation task and a 6x speedup on the BERT task without a significant loss in model performance. HetSeq does not show performance gains on the MNIST task because of the small number of training samples.}
\vspace{-.3cm}
\resizebox{\textwidth}{!}{%
\begin{tabular}{@{}clrrrrrrllrr@{}}
\toprule
& \textbf{nodes} & \textbf{GPUs} & \textbf{epochs} & \textbf{max tokens} & \textbf{steps per GPU}   & \textbf{avg. step time} & \textbf{training time}   & \multicolumn{2}{l}{\textbf{BLEU4 score}}                          & \textbf{expansion} & \textbf{speedup}       \\ \midrule
\multirow{6}{*}{\rotatebox[]{90}{Translator}} & 1     & 4    & 128    & 4,096       & 260,000 & 0.62 s    & 49.47 hr         & \multicolumn{2}{l}{25.09, 56.8/30.8/18.9/12.0}        & 1.00      & 1.00          \\ 
& 2 (hom)     & 8    & 128    & 8,192       & 130,000 & 0.90 s             & 34.23 hr        & \multicolumn{2}{l}{25.16, 57.1/30.9/18.9/12.0} & 0.73     & 1.45          \\ 
& 2 (het)     & 8    & 128    & 8,192       & 130,000 & 0.90 s             & 34.81 hr         & \multicolumn{2}{l}{25.57, 57.2/31.2/19.3/12.4} & 0.71      & 1.42          \\ 
& 4 (hom)   & 16   & 128    & 16,384      & 65,000  & 0.94 s           & 18.12 hr         & \multicolumn{2}{l}{24.98, 56.7/30.5/18.7/12.0}          & 0.68      & 2.73          \\ 
& 4 (het)   & 16   & 128    & 16,384      & 65,000  & 0.94 s             & 18.74 hr         & \multicolumn{2}{l}{25.19, 57.1/30.9/18.9/12.1}          & 0.66      & 2.64          \\ 
& 8 (het)    & 32   & 128    & 32,768      & 32,500  & 0.98 s             & 10.3 hr & \multicolumn{2}{l}{18.74, 52.2/24.4/13.2/7.5}           & 0.60      & 4.80 \\ \bottomrule
\toprule
& \textbf{nodes} & \textbf{GPUs} & \textbf{epochs} & \textbf{batch-size} & \textbf{steps per GPU}   & \textbf{avg. step time} & \textbf{training time}  & \multicolumn{2}{l}{\textbf{training loss}}  & \textbf{expansion} & \textbf{speedup}       \\ \midrule
\multirow{6}{*}{\rotatebox[]{90}{BERT}} &  1     & 4    & 5      & 128        & 267,139 & 2.60 s     & 7.19 d          & \multicolumn{2}{l}{0.026} & 1.00      & 1.00          \\
& 2 (hom)     & 8    & 5      & 256        & 133,570 & 2.69 s              & 4.19 d          & \multicolumn{2}{l}{0.028}          & 0.86      & 1.72          \\ 
& 2 (het)     & 8    & 5      & 256        & 133,570 & 2.74 s              & 4.26 d          & \multicolumn{2}{l}{0.028}          & 0.85      & 1.69          \\ 
& 4 (hom)     & 16   & 5      & 512        & 66,785  & 2.79 s              & 2.23 d & \multicolumn{2}{l}{0.031}          & 0.81      & 3.22 \\
& 4 (het)     & 16   & 5      & 512        & 66,785  & 2.81 s              & 2.19 d  & \multicolumn{2}{l}{0.031}          & 0.82      & 3.28 \\
& 8 (het)     & 32   & 5      & 1024        & 33,393  & 3.13 s              & 1.21 d  & \multicolumn{2}{l}{0.055}          & 0.74      & 5.94 \\
\bottomrule
\toprule
& \textbf{nodes} & \textbf{GPUs} & \textbf{epochs} & \textbf{batch-size} & \textbf{steps per GPU}   & \textbf{avg. step time} & \textbf{training time} & \textbf{test loss}  & \textbf{test accuracy}  & \textbf{expansion} & \textbf{speedup}       \\ \midrule
\multirow{6}{*}{\rotatebox[]{90}{MNIST}} & 1     & 4    & 20      & 256        & 47,00 & 0.00612 s  & 87.6s   & 0.0005          & 0.9918 & 1.00 & 1.00 \\
& 2 (hom)     & 8    & 20      & 512        & 2,360 & 0.01364 s    & 87.0s    & 0.0004         & 0.9911   & 0.50 & 1.01       \\ 
& 2 (het)     & 8    & 20      & 512        & 2,360 & 0.01406 s     & 87.7s   & 0.0004          & 0.9912    & 0.50 & 1.00      \\ 
& 4 (hom)     & 16   & 20      & 1024        & 1,180  & 0.01490 s   & 72.1s   & 0.0004   & 0.9912   & 0.30 & 1.21    \\
& 4 (het)     & 16   & 20      & 1,024        & 1,180  &  0.01473 s     & 72.6s   & 0.0005  & 0.9916  & 0.30 & 1.21    \\
& 8 (het)     & 32   & 20      & 2,048        & 600  & 0.01745 s     & 80.6s      & 0.0005  & 0.9909  & 0.14 & 1.09 \\
\bottomrule
\end{tabular}%
}
\label{exp}
\vspace{-.2cm}
\end{table*}

\subsection{Transformer Translation Model}
We evaluate HetSeq using the base Transformer model~\cite{vaswani2017attention} on the WMT 2014 English-to-German translation task (En-to-De). The model has $6$ encoder layers and $6$ decoder layers. The size of the word embedding (\ie, hidden state) is 512 and the number of heads is 8. Dropout rate is set to $0.1$ and we use the label-smoothed cross entropy loss function with $\epsilon = 0.1$. In total, this model has about 65 million parameters. We use the Adam optimizer with $\beta_{1} = 0.9$, $\beta_{2} = 0.98$ and $\epsilon = 10^{-9}$.

\paragraph{Results.}
Table~\ref{exp} shows the results of the transformer experiments on 2014 English-to-German dataset. We record the total training time and BLEU4 score (for the average and 1-, 2-, 3-, and 4-grams), which is a standard evaluation metric for language translation, for each experiment. %We find that training time is dramatically improved. 

The heterogeneous configurations show that speedup scales at about one-half the linear rate, which can certainly be improved with further development. As the number of nodes increases from 1 to 2 (\ie, 4 to 8 GPUs), the training time is sped up by a factor of 1.42. As more nodes are added, the performance improvement (\ie, expansion) decreases to 0.6 over 8 heterogeneous nodes, but a nearly 5x speed up is achieved. This allows for the transformer to be trained in only 10 hours. %We also measure HetSeq's efficiency with the expansion metric, calculated as the ratio of the speedup to the number of GPUs. Expansion is reasonable at 0.6 in the 8 node configuration.

%I rolled expansion into the earlier sentence. Does this make sense?

%this is too much detail for prose - they can read the table for this.

Critically, the performance of the heterogeneous configurations are rather similar to the performance of the homogeneous configurations. These results indicate that the performance of HetSeq does not deteriorate significantly in the presence of heterogeneous infrastructure - at least as compared to homogeneous infrastructure. % Although we had access to eight equivalent nodes, they existed as shared resources; so unfortunately, we were not able to capture eight equivalent nodes for an eight-node homogeneous comparison.

We also find that the BLEU score does vary for each task. This is because the different experiments are conducted with different batch sizes and the same optimizer and learning rate scheduling set ups. Although the 2 node configuration resulted the best performance over the same number of steps, the model performance was relatively consistent across configurations.

In summary, we show that HetSeq can speed up training on the transformer model on heterogeneous infrastructure without sacrificing model performance.

%In each experiment, we measure the time of each training step, total training time and BLEU score of each trained model. The expansion and speed up calculation is all based on total training time. From table~\ref{transformer}, one can observe that, by employing our method, the training speed increases. The speed up gets $1.42$ when using $2$ nodes, gets $2.64$ when using $4$ nodes, and gets $4.89$ when using $8$ nodes. It's interesting that scaling from $1$ node to $2$ nodes, the average time of each step increases $0.28$s compared to original $0.62$s. In comparison, scaling from $2$ nodes to $4$ nodes or from $4$ nodes to $8$ nodes, the averaging step time increases just $0.04$s. So we want to significantly reduce the training time, we should use more GPUs to reduce the impact of initial cost. 

\subsection{BERT Language Model}
The BERT language model~\cite{devlin2018bert} masks some of the words in a sentence and tries to infer identities of the masked words using information from the unmasked words. We evaluate HetSeq using the base BERT language model trained on the Wikipedia corpus. %Originally, the BERT language model was also trained on BookCorpus, but this dataset is no longer publicly available. 
We train base BERT using the Adam optimizer with $\beta_{1} = 0.9$, $\beta_{2} = 0.999$ and $\epsilon = 10^{-8}$. We use the standard linear decay learning rate scheduler with maximum learning rate $= 0.0001$. In total, the model has over 1 billion parameters. In all cases, we use the first 10K steps as the warm up and 1 million steps total over all GPUs.

\paragraph{Results.}
Runtime and training loss are described in Tab.~\ref{exp}. We show significant speedup as the number of GPUs and nodes increases. Despite fewer steps on each GPU, the total training loss performance is maintained as the number of nodes increases. 

Again we find that heterogeneous and homogeneous configurations achieve roughly the same speedup. However, it is important to note that in the both language model and transformer experiments these runtime results are not apples-to-apples comparisons. For example, the P100s included in the heterogeneous configuration are faster than the Xps used in the homogeneous experiments. We do not intend the current work to be a full system analysis, but rather provide these runtime benchmarks as as evidence of HetSeq's scalability.

In summary, HetSeq is able to reduce the training time for the BERT language model on a single node from seven days to about one day on heterogeneous infrastructure.

\subsection{MNIST Image Classification Model}
To show the extensibility of HetSeq to other models, we also implement the image classification model from PyTorch and evaluate it within HetSeq. This model contains two layers of a convolutional neural network follows by a flat fully connected layer to perform image classification with a cross entropy loss~\cite{simard2003best}. We use the Adam optimizer with $\beta_{1} = 0.9$, $\beta_{2} = 0.999$ and $\epsilon = 10^{-8}$. The starting learning rate is set to $1.01$ for all the experiments.

\paragraph{Results.} Runtime and accuracy on test set are described in Tab.~\ref{exp}. Compared to the translation and language models, the image classifier does not show good scaling with HetSeq. But this is merely because the model and dataset is so small. Specifically, MNIST has only $60,000$ training examples which we load directly into memory. Furthermore, the model requires fewer than $5,000$ training steps on single GPU, which is only a small fraction of the actual training time. Nevertheless, this task is useful because it shows how to extend HetSeq to other kinds of models. We expect that training over much larger datasets will more clearly reveal the benefits of HetSeq.

%As for test accuracy, the results keep similar and high performance across all the settings. Furthermore, the avg. step time shows the potential scalability with large model and datasets (\eg, ImageNet or COCO) on similar tasks. 

%\input{thesis_proposal}

\section{Discussion}
\label{sec:discussion}
The present work describes HetSeq, a publicly available deep learning platform adapted from PyTorch that enables distributed GPU training on heterogeneous infrastructure. HetSeq works by duplicating and distributed the model architecture to each GPU, which is its own process having its own optimizer, learning rate scheduler, data loader, etc. Each GPU communicates the loss and gradients while performs parameter update individually. Experiments on the transformer and BERT language model show that HetSeq can achieve reasonable speedup even over heterogeneous infrastructure.

HetSeq can be further extended to incorporate other deep learning models in natural language processing, computer vision, and elsewhere. Future plans include adapting ongoing research in distributed optimization~\cite{you2019large} to further improve training performance on heterogeneous infrastructure.

\section{Acknowledgements}
We thank Satyaki Sikdar for his help preparing this paper. This work is funded by the US Army Research Office (W911NF-17-1-0448) and the US Defense Advanced Research Projects Agency (DARPA W911NF-17-C-0094).

%\begin{small}
%\bibliography{references}
%\end{small}

\end{document}